\documentclass[10pt,epsfig]{article}

\usepackage{multicol}
\usepackage{graphicx}
\usepackage{booktabs}
\usepackage{amssymb,bm,mathrsfs,bbm,amscd}
\usepackage[tbtags]{amsmath}
\usepackage{lastpage}
\usepackage{lscape}
\usepackage[rotateright]{rotating}

\begin{document}
\title{Forward-backward emission of target evaporated  evaporated fragments at high energy nucleus-nucleus collisions
\thanks{Submitted to Chin. Phys. C}}

\author{Zhi Zhang, Tian-Li Ma,Dong-Hai Zhang\thanks{Corresponding author. Tel: +863572051347; fax: +863572051347. E-mail address:zhangdh@dns.sxnu.edu.cn}\\
Institute of Modern Physics, Shanxi Normal University, Linfen 041004, China}

\date{}
\maketitle

\begin{center}
\begin{minipage}{140mm}
\vskip 0.4in
\begin{center}{\bf Abstract}\end{center}
{The multiplicity distribution, multiplicity moment, scaled variance, entropy and reduced entropy of target evaporated fragment emitted in forward and backward hemispheres in 12 A GeV $^{4}$He, 3.7 A GeV $^{16}$O, 60 A GeV $^{16}$O, 1.7 A GeV $^{84}$Kr and 10.7 A GeV $^{197}$Au induced emulsion heavy targets (AgBr) interactions are investigated. It is found that the multiplicity distribution of target evaporated fragments emitted in forward and backward hemispheres can be fitted by a Gaussian distribution. The multiplicity moments of target evaporated particles emitted in forward and backward hemispheres increase with the order of the moment {\em q}, and second-order multiplicity moment is energy independent over the entire energy for all the interactions in the forward and backward hemisphere respectively. The scaled variance, a direct measure of multiplicity fluctuations, is close to one for all the interactions which may be said that there is a feeble correlation among the produced particles. The entropy of target evaporated fragments emitted in forward and backward hemispheres are the same within experimental errors, respectively.}\\

\end{minipage}
\end{center}

\vskip 0.4in
\baselineskip 0.2in
\section{Introduction}

In high energy nucleus-nucleus collisions, the multiplicity is an important variable which can help to test different phenomenological and theoretical models and to understand the mechanism of multi-particle production. According to the participant-spectator model\cite{bowm}, projectile fragments, target fragments and produced particles, three different types of secondary particles are produced in high energy nucleus-nucleus collisions. The produced particles are single-charged relativistic particles having velocity $v\geq0.7$\,c. Most of them belong to pions contaminated with small proportions of fast protons and K mesons. The multiplicity of produced particles has been studied exclusively because they carry out the essential information of the interaction mechanism. The projectile fragments are the decayed particles of the excited projectile residues through evaporation of neutrons and light nuclei, it has been speculated that the decay of highly excited nuclear matter system may carry information about the equation of state and the liquid-gas phase transition of low-density nuclear matter. The target fragments include target recoiled protons and target evaporated fragments. The target recoiled protons are formed due to fast target protons of energy ranging up to $400$ MeV. The target evaporated particles are of low-energy ($<30$ MeV) singly or multiply charged fragments. In emulsion terminology\cite{powe}, the target recoiled particles are referred to as "grey track particles" and the target evaporated particles are referred to as "black track particles". According to the cascade evaporation model\cite{powe}, the grey track particles are emitted from the nucleus very soon after the instant of impact, leaving the hot residual nucleus in an excited state. The emission of black track particles from this state takes place relatively slowly. In the rest system of the target nucleus, the emission of evaporated particles is assumed to be isotropic in whole phase space. Much attention has been paid to investigate the production of grey track particles because it is believed to be a quantitative probe of the intranuclear cascading collisions. For the study of the production of black particles, a little attention is paid because these particles are emitted at a late stage of nuclear interactions. However, the study of target evaporated particles is also important because they are expected to remember the history of the interactions. Hence, the emission of these particles may also be of physical interest.

The emission of target evaporated fragments is isotropic in the rest system of target nucleus according to the cascade evaporation model, but attributed to the electromagnetic field from projectile, this isotropic emission property may be distorted. So it is necessary to compare the emission properties of target evaporated fragments in forward hemisphere(emission angle $\theta_{Lab}\leq90^{\circ}$) and that in backward hemisphere(emission angle $\theta_{Lab}>90^{\circ}$). If the nature of multiplicity distribution is the forward and backward hemispheres is similar, then isotropic target evaporated fragments emission may be assumed. In contract, if the multiplicity distributions in the two hemispheres are different, then the emission mechanism of target evaporated fragments in the forward and backward hemispheres may be different. This can be interesting and deserves the attention of the physics community.

Ghosh et al.\cite{ghosh} has studied the emission properties of target evaporated fragments in forward and backward hemispheres produced in 3.7 A GeV $^{12}$C-AgBr, 14.5 A GeV $^{28}$Si-AgBr, 60 A GeV $^{16}$O-AgBr and 200 A GeV $^{32}$S-AgBr interactions, the difference in multiplicity distribution is found in the forward and backward hemispheres. In our recent paper\cite{zhang} the forward-backward correlations for target fragments emitted in 150 A MeV $^{4}$He-AgBr, 290 A MeV $^{12}$C-AgBr, 400 A MeV $^{12}$C-AgBr, 400 A MeV $^{20}$Ne-AgBr and 500 A MeV $^{56}$Fe-AgBr interactions are studied.

In this paper, the production of target evaporated fragments emitted from 12 A GeV $^{4}$He-AgBr, 3.7 A GeV $^{16}$O-AgBr, 60 A GeV $^{16}$O-AgBr, 1.7 A GeV $^{84}$Kr-AgBr and 10.7 A GeV $^{197}$Au-AgBr interactions are investigated. The multiplicity distribution, multiplicity moments, the scaled variance, entropy and reduced entropy of target evaporated fragments emitted in forward and backward hemispheres are discussed respectively.

\section{Experimental details}

Five nuclear emulsion stacks, provided by EMU01 Collaboration, are used in present investigation. The stacks were exposed horizontally by 12 A GeV $^{4}$He, 3.7 A GeV $^{16}$O, 60 A GeV $^{16}$O, 1.7 A GeV $^{84}$Kr and 10.7 A GeV $^{197}$Au. BA2000 and XSJ-2 microscopes with a $100\times$ oil immersion objective and $10\times$ ocular lenses were used to scan the plates. The tracks were picked up at a distance of $5$\,mm from the edge of the plates and they were carefully followed until they either interacted with emulsion nuclei or escaped from the plates. Interactions were within $30$\,$\mu$\,m from the top or bottom surface of the emulsion plates were not considered for final analysis. In each interaction all of the secondaries were recorded which include shower particles, target recoiled protons, target evaporated fragments and projectile fragments. Details of track scanning and track classification can be found in our paper\cite{zhang}.

The nuclear emulsion is composed of a homogeneous mixture of nucleus H, C, N, O, S, I, Br, and Ag, and major composition is H, C, N, O, Br, and Ag. According to the value of $n_{h}$(multiplicity of the target recoiled protons and evaporated fragments) the interactions are divided into following three groups. Events with $n_{h}\leq1$ are due to interactions with H target and peripheral interactions with CNO and AgBr targets, events with $2\leq{n_{h}}\leq7$ are due to interactions with CNO targets and peripheral interactions with AgBr targets, and events with $n_{h}\geq8$ definitely belong to interactions with AgBr targets.

To ensure the targets in nuclear emulsion are silver or bromine nuclei, we have chosen only the events with at least eight heavy ionizing track particles.

\section{Results and discussion}

The general characteristics of 12 A GeV $^{4}$He, 3.7 A GeV $^{16}$O, 60 A GeV $^{16}$O, 1.7 A GeV $^{84}$Kr and 10.7 A GeV $^{197}$Au induced AgBr interactions including event statistics, average multiplicity of the target evaporated fragments emitted in forward hemisphere($<n_{b}^{f}>$), backward hemisphere($<n_{b}^{b}>$) and whole space($<n_{b}>$), are presented in Table\,1. It is found that the target evaporated fragments emitted in forward hemisphere is greater than that in backward hemisphere.

\begin{table}
\begin{center}
Table 1.The average multiplicity of target evaporated fragments in the forward and backward hemispheres for different nucleus-AgBr interactions.\\
\begin{tabular}{lcccc}\hline
 Beam                    & events  & $<n_{b}^{f}>$ & $<n_{b}^{b}>$ & $<n_{b}>$  \\\hline
12 A GeV $^{4}$He     & $2975$  & $4.62\pm0.05$ & $3.65\pm0.04$ & $8.26\pm0.08$ \\
3.7 A GeV $^{16}$O    & $927$   & $6.04\pm0.11$ & $4.47\pm0.08$ & $10.50\pm0.16$ \\
60 A GeV $^{16}$O     & $521$   & $5.98\pm0.12$ & $5.12\pm0.11$ & $11.05\pm0.19$ \\
1.7 A GeV $^{84}$Kr   & $229$   & $6.27\pm0.21$ & $4.07\pm0.15$ & $10.34\pm0.29$ \\
10.7 A GeV $^{197}$Au & $619$   & $5.01\pm0.11$ & $3.66\pm0.09$ & $8.67\pm0.17$ \\\hline
\end{tabular}
\end{center}
\end{table}

Figures 1 to 5 show the multiplicity distribution of target evaporated fragment emitted in forward hemisphere, backward hemisphere and whole space. It is found that the distribution can be well fitted by a Gaussian distribution for 3.7 A GeV $^{16}$O, 60 A GeV $^{16}$O, 1.7 A GeV $^{84}$Kr and 10.7 A GeV $^{197}$Au induced AgBr interactions, but for 12 A GeV $^{4}$He induced AgBr target interactions the distributions can be fitted by the superposition of two Gaussian distribution. The Gaussian fitting parameters (mean value and error) and $\chi^{2}$/DOF are presented in Table\,2, where DOF means the degree of freedom of simulation. For comparison the results in Ref.\cite{ghosh} are also included in the table. It is found that the fitting parameters are different between the forward and backward hemispheres for all the interactions, the mean values and errors of Gaussian distributions in forward hemisphere are greater than that in backward hemisphere. The difference in the nature of multiplicity distribution between the two hemispheres may be attributed to the fact that the mechanism of the target fragmentation process is different in the forward and backward hemispheres. Based on the cascade evaporation model\cite{powe} the emission of the target evaporated fragments should be isotropic in the laboratory frame, but attributed to the electromagnetic field from projectile, the emission of target evaporated fragments is close to $\theta_{Lab}\approx90^{\circ}$ symmetric and the emission probability in forward hemisphere is greater than that in backward hemisphere.  According to the model proposed by Stocker et al.\cite{stock} using three-dimensional nuclear fluid dynamics, the emission of target fragments in backward hemisphere can be explained with the help of the side splash phenomenon. In a nucleus-nucleus collision, a head shock zone may be developed during the dividing phase of the projectile nucleus with the target. A strongly compressed and highly excited projectile-like object continues to interpenetrate the target with supersonic velocity and may push the matter sideward. This results in the generation of shock waves that give rise to particle evaporation in the backward directions. At intermediate impact parameters the highly inelastic bounce-off appears, where the large compression potential leads to the sidewards deflection of the projectile, which then explodes. A large collective transverse momentum transfer to the target leads to azimuthally asymmetric fragment distribution.

\begin{figure}[htbp]
\begin{center}
\includegraphics[width=0.67\linewidth]{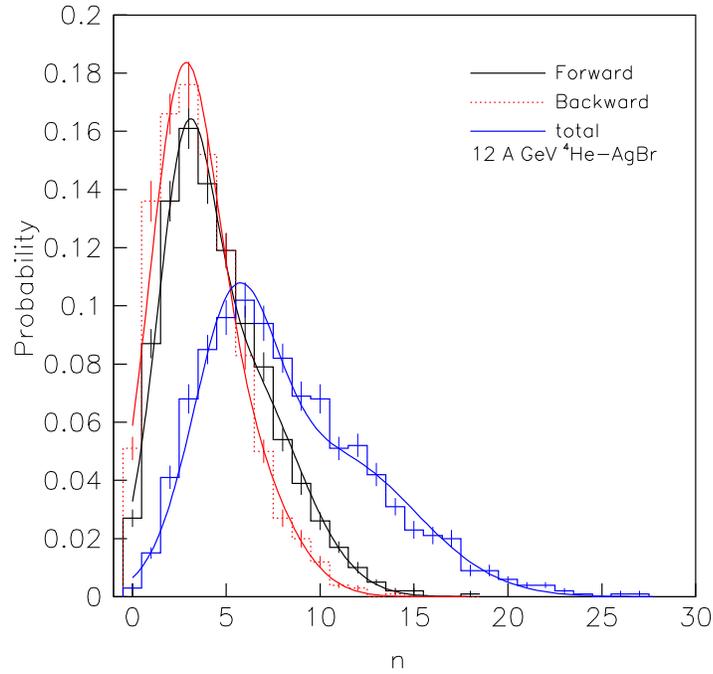}
\caption{(Color online) Multiplicity distributions of target evaporated fragments emitted in 12 A GeV $^{4}$He-AgBr interactions, the smooth curves are the results from the Gaussian fitting.}
\end{center}
\end{figure}

\begin{figure}[htbp]
\begin{center}
\includegraphics[width=0.67\linewidth]{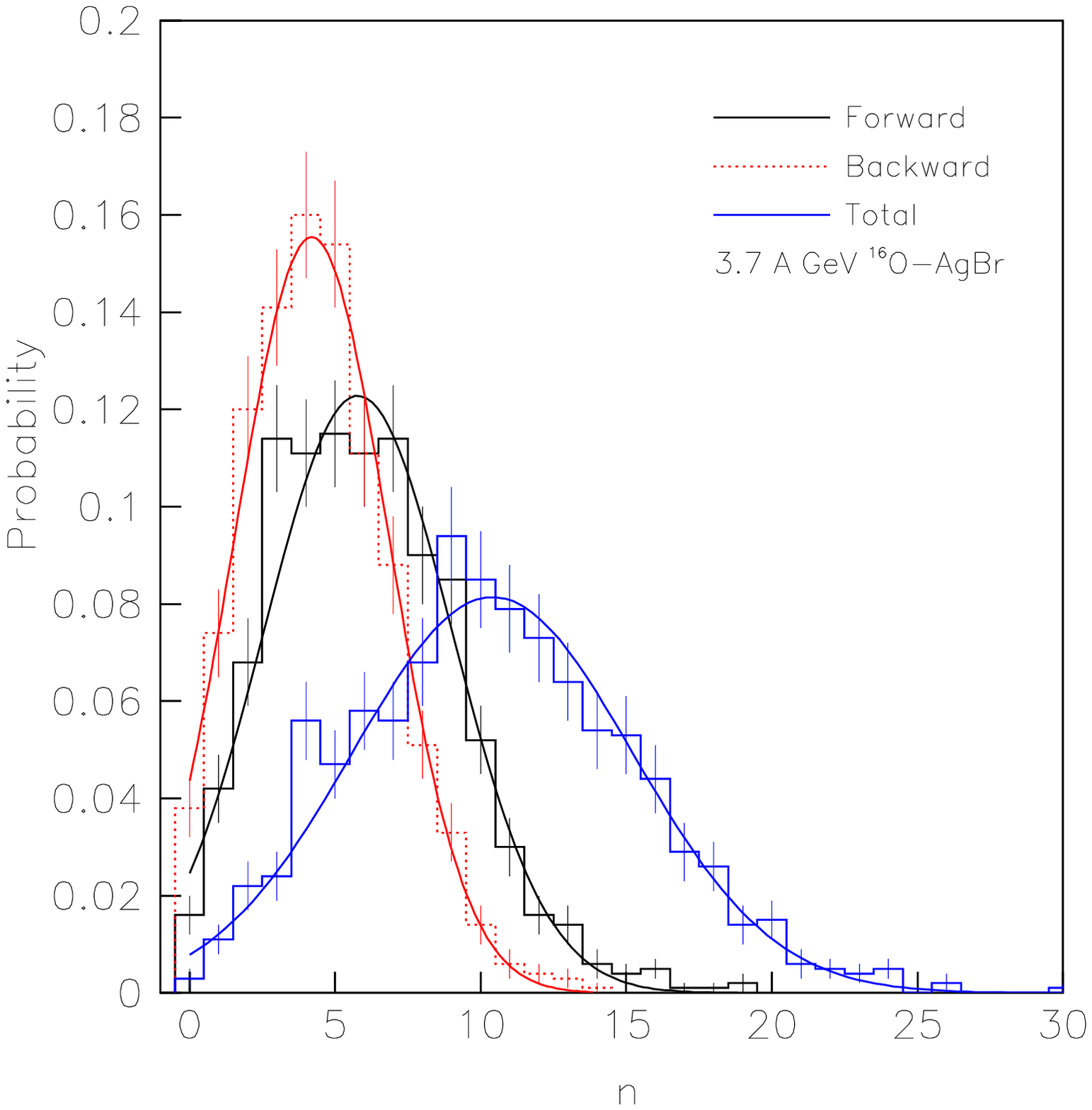}
\caption{(Color online) Multiplicity distributions of target evaporated fragments emitted in 3.7 A GeV $^{16}$O-AgBr interactions, the smooth curves are the results from the Gaussian fitting.}
\end{center}
\end{figure}

\begin{figure}[htbp]
\begin{center}
\includegraphics[width=0.67\linewidth]{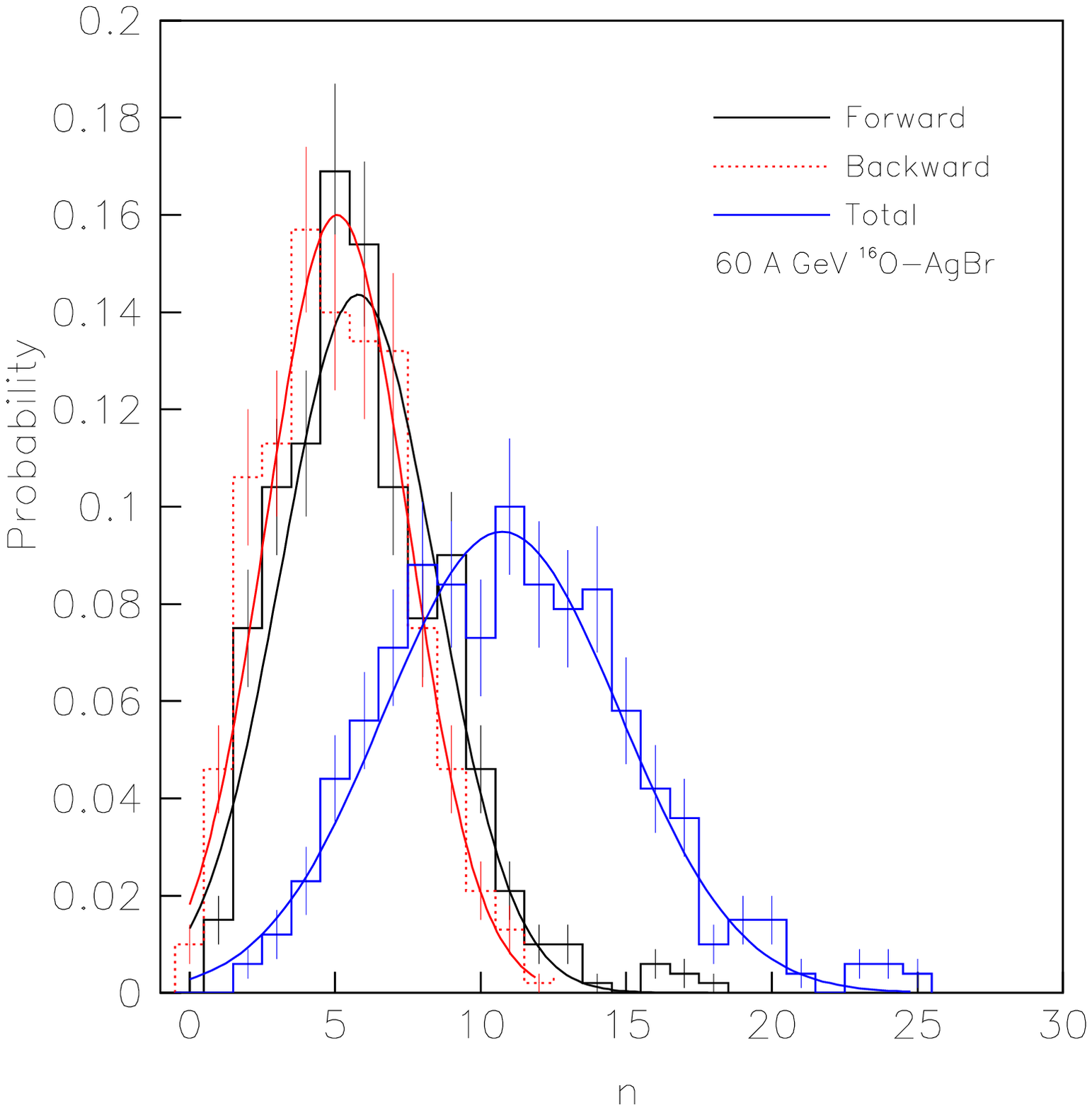}
\caption{(Color online) Multiplicity distributions of target evaporated fragments emitted in 60 A GeV $^{16}$O-AgBr interactions, the smooth curves are the results from the Gaussian fitting.}
\end{center}
\end{figure}

\begin{figure}[htbp]
\begin{center}
\includegraphics[width=0.67\linewidth]{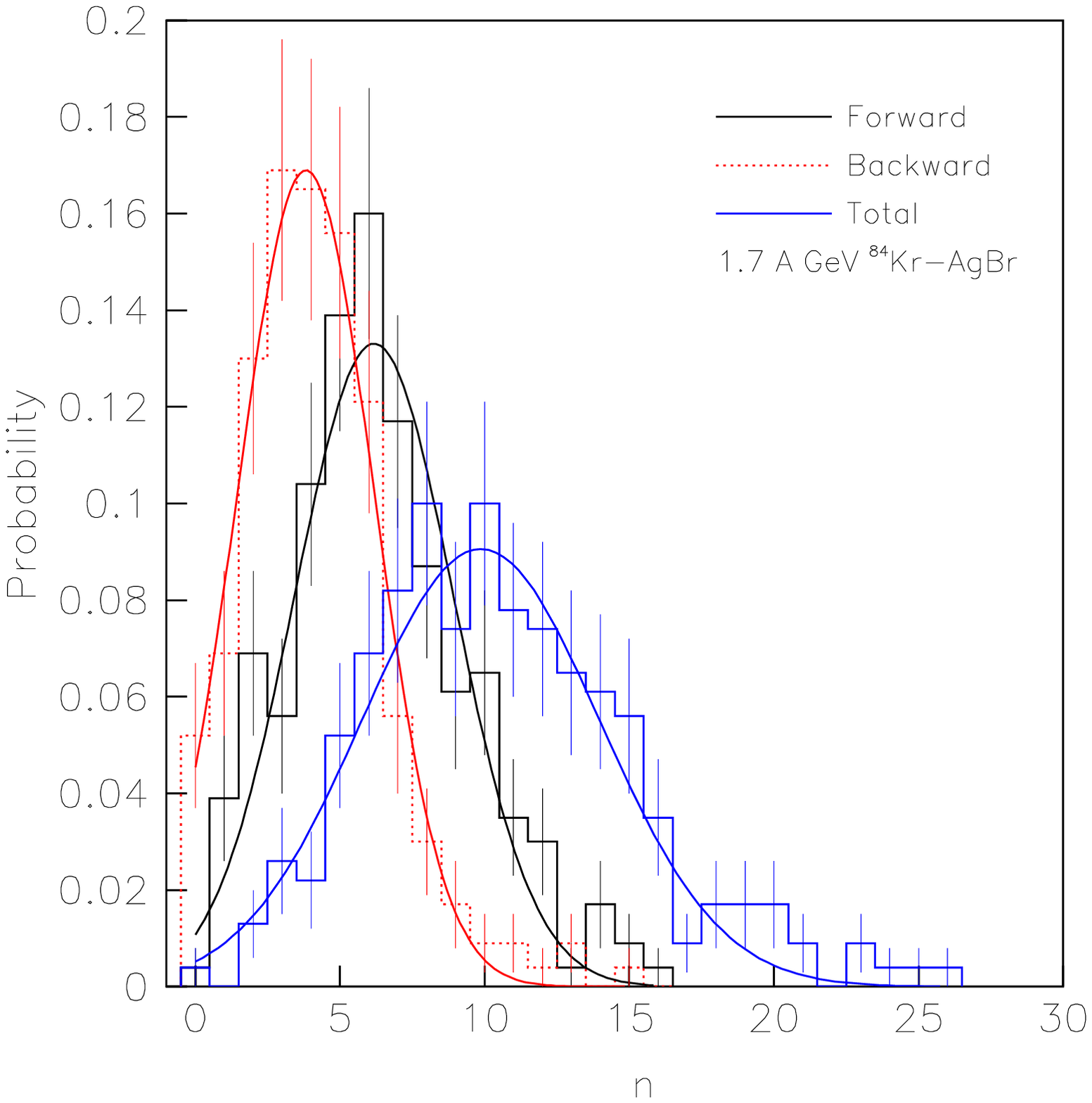}
\caption{(Color online) Multiplicity distributions of target evaporated fragments emitted in 1.7 A GeV $^{84}$Kr-AgBr interactions, the smooth curves are the results from the Gaussian fitting.}
\end{center}
\end{figure}

\begin{figure}[htbp]
\begin{center}
\includegraphics[width=0.67\linewidth]{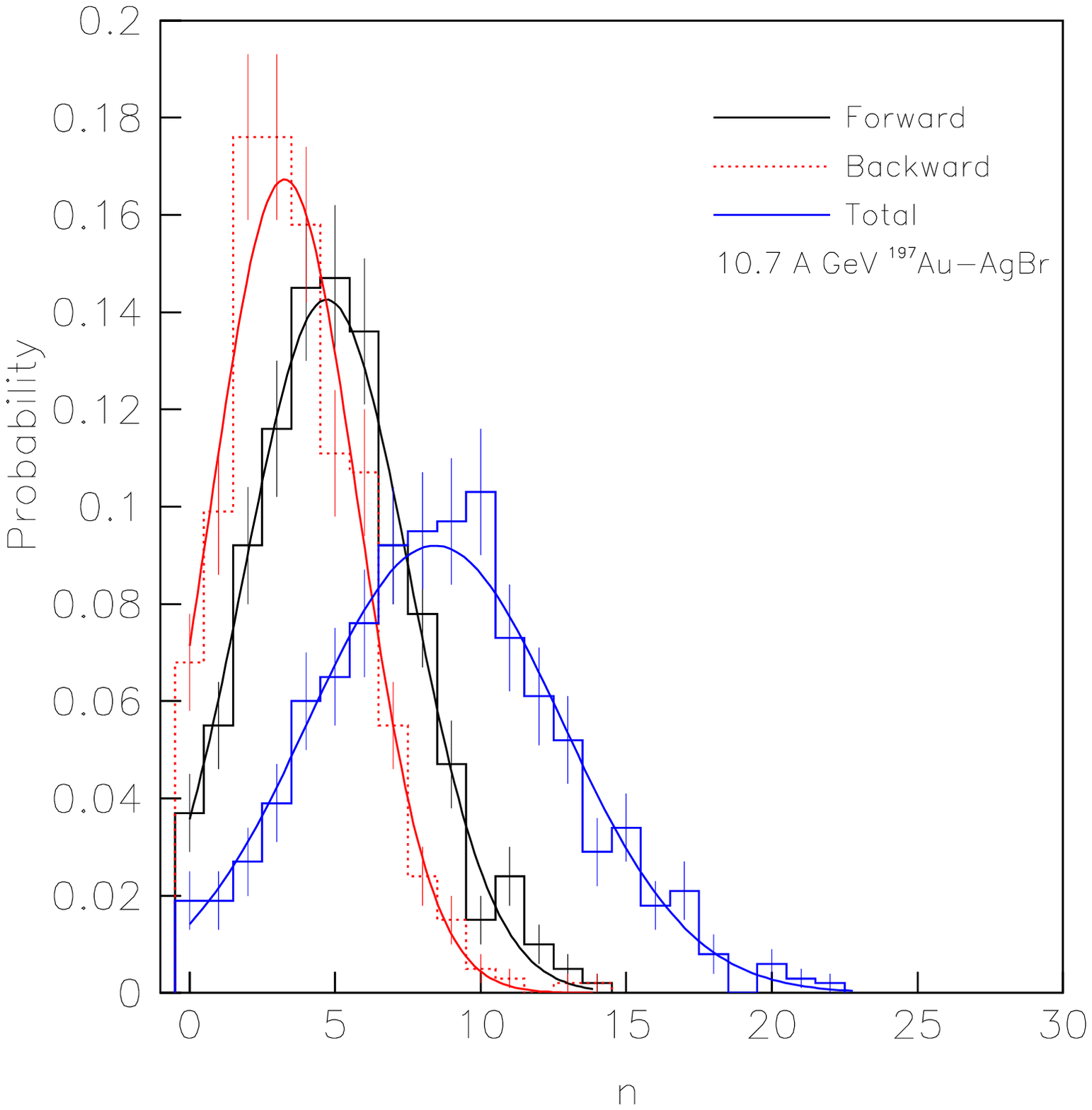}
\caption{(Color online) Multiplicity distributions of target evaporated fragments emitted in 10.7 A GeV $^{197}$Au-AgBr interactions, the smooth curves are the results from the Gaussian fitting.}
\end{center}
\end{figure}

Equivalently, the multiplicity distribution can be studied using the moments of the distribution, which is given by
\begin{equation}
C_{q}=\frac{<n^{q}>}{<n>^{q}}=\frac{\sum_{n}n^{q}P_{n}}{(\sum_{n}nP_{n})^{q}}
\end{equation}
where {\em q} is a positive integer called the order of the moment and $P_{n}$ the probability of producing or emitting {\em n} particles. Here
\begin{equation}
<n^{q}>=\sum_{n}n^{q}\frac{\sigma_{n}}{\sigma_{inel}}
\end{equation}
where $\sigma_{n}$ is the partial cross section for producing or emitting a state of multiplicity {\em n}, $\sigma_{inel}$ is the total inelastic cross-section and $<n>$ is the average multiplicity.

The multiplicity moments, sometimes called the reduced {\em C} moments, can be used to describe the properties of multiplicity distributions, e.g. as a function of the projectile energy in the center-of-mass frame or in the laboratory frame. In practice, only the first few moments can be calculated with reasonable accuracy due to the limited statistics. Table\,3 presents the multiplicity moments of target evaporated fragments emitted in 12 A GeV $^{4}$He, 3.7 A GeV $^{16}$O, 60 A GeV $^{16}$O, 1.7 A GeV $^{84}$Kr and 10.7 A GeV $^{197}$Au induced AgBr interactions, the corresponding results in Ref.\cite{ghosh} are also included. In both the hemispheres, for all the interactions, multiplicity moments increase with the order of the moment {\em q}, and the second order multiplicity moments is energy independent over the entire energy range in the forward and backward hemispheres respectively. In the backward hemisphere, multiplicity moments up to the third order are also energy independent for all the interaction. The energy-independent behavior of multiplicity moments may hint at the existence of KNO scaling, which is an energy independent scaling law of multiplicity distribution proposed by Koba, Nielsen and Olesen\cite{koba}.

\begin{table}
\begin{center}
Table 2. The Gaussian fitting parameters of target evaporated fragment multiplicity distributions in the forward and backward hemispheres for different type of nucleus-AgBr interactions.\\
\begin{small}
\begin{tabular}{lcccccccccc}\hline
Beam             &\multicolumn{3}{c}{forward hemisphere} &\multicolumn{3}{c}{backward hemisphere} &\multicolumn{3}{c}{whole space} & Reference \\
                     & mean value    & error & $\chi^{2}$ & mean value & error  & $\chi^{2}$ & mean value & error  & $\chi^{2}$ &    \\\hline
12 A GeV He   & $2.80\pm0.11$ & $1.49\pm0.13$ & $1.57$ & $2.59\pm0.16$ & $1.76\pm0.18$ & $2.97$ & $5.38\pm0.15$ & $2.19\pm0.11$ & $2.72$ & This work\\
                 & $5.68\pm0.41$ & $3.03\pm0.15$ &        & $4.94\pm1.18$ & $2.69\pm0.30$ &        & $10.69\pm0.37$& $4.38\pm0.16$ &        & This work\\
3.7 A GeV O   & $5.79\pm0.12$ & $3.23\pm0.10$ & $1.47$ & $4.20\pm0.10$ & $2.64\pm0.09$ & $0.60$ & $10.40\pm0.17$& $4.81\pm0.14$ & $1.17$ & This work\\
60 A GeV O    & $5.79\pm0.13$ & $2.65\pm0.11$ & $2.59$ & $5.08\pm0.11$ & $2.43\pm0.08$ & $1.56 $& $10.74\pm0.19$& $4.05\pm0.15$ & $1.31$ & This work\\
1.7 A GeV Kr  & $6.18\pm0.20$ & $2.76\pm0.16$ & $1.47$ & $3.82\pm0.18$ & $2.36\pm0.18$ & $0.64$ & $9.87\pm0.30$ & $4.12\pm0.27$ & $0.83$ & This work\\
10.7 A GeV Au & $4.72\pm0.13$ & $2.84\pm0.12$ & $1.07$ & $3.27\pm0.12$ & $2.50\pm0.11$ & $1.03$ & $8.44\pm0.19$ & $4.37\pm0.20$ & $0.60$ & This work\\
3.7 A GeV C   & $5.00$        & $3.26$        & $0.51$ & $4.57$        & $2.80$        & $0.48$ &               &               &        & [3] \\
14.5 A GeV Si & $5.27$        & $2.13$        & $0.32$ & $4.27$        & $1.93$        & $0.87$ &               &               &        & [3]\\
60 A GeV O    & $7.19$        & $2.93$        & $0.54$ & $4.53$        & $2.99$        & $0.29$ &               &               &        & [3] \\
200 A GeV S   & $4.78$        & $4.46$        & $0.68$ & $3.89$        & $2.80$        & $0.61$ &               &               &        & [3]\\\hline
\end{tabular}
\end{small}
\end{center}
\end{table}

\begin{table}
\begin{center}
Table 3. The values of the multiplicity moments of target evaporated fragments in the forward and backward hemispheres for different type of nucleus-AgBr interactions.\\
\begin{small}
\begin{tabular}{lccccccc}\hline
Beam    &\multicolumn{3}{c}{forward hemisphere} &\multicolumn{3}{c}{backward hemisphere} & Reference  \\
                 & $C_{2}$       & $C_{3}$       & $C_{4}$       & $C_{2}$       & $C_{3}$       & $C_{4}$       &    \\\hline
12 A GeV He   & $1.38\pm0.06$ & $2.35\pm0.17$ & $4.67\pm0.24$ & $1.41\pm0.06$ & $2.43\pm0.17$ & $4.87\pm0.29$ & This work\\
3.7 A GeV O   & $1.28\pm0.09$ & $1.94\pm0.21$ & $3.35\pm0.25$ & $1.31\pm0.09$ & $2.02\pm0.22$ & $3.52\pm0.32$ & This work\\
60 A GeV O    & $1.22\pm0.10$ & $1.74\pm0.23$ & $2.87\pm0.25$ & $1.23\pm0.10$ & $1.72\pm0.21$ & $2.65\pm0.26$ & This work\\
1.7 A GeV Kr  & $1.24\pm0.16$ & $1.77\pm0.34$ & $2.84\pm0.40$ & $1.36\pm0.22$ & $2.30\pm0.60$ & $4.66\pm0.84$ & This work\\
10.7 A GeV Au & $1.29\pm0.11$ & $1.94\pm0.24$ & $3.25\pm0.34$ & $1.39\pm0.13$ & $2.32\pm0.35$ & $4.50\pm0.60$ & This work\\
3.7 A GeV C   & $1.32\pm0.09$ & $2.12\pm0.12$ & $3.90\pm0.21$ & $1.26\pm0.03$ & $1.86\pm0.12$ & $2.95\pm0.32$ & [3] \\
14.5 A GeV Si & $1.20\pm0.07$ & $1.63\pm0.14$ & $2.45\pm0.28$ & $1.23\pm0.04$ & $1.79\pm0.13$ & $2.98\pm0.47$ & [3]\\
60 A GeV O    & $1.19\pm0.08$ & $1.63\pm0.15$ & $2.51\pm0.37$ & $1.24\pm0.05$ & $1.83\pm0.15$ & $3.03\pm0.63$ & [3] \\
200 A GeV S   & $1.30\pm0.05$ & $1.94\pm0.18$ & $3.16\pm0.46$ & $1.30\pm0.08$ & $2.06\pm0.21$ & $3.92\pm0.81$ & [3]\\\hline
\end{tabular}
\end{small}
\end{center}
\end{table}

It is known that the probability of multiplicity can be used to evaluate the entropy of the produced particles. The entropy of the produced particles can be calculated using the formula defined by Wehrl\cite{wehrl} as
\begin{equation}
S=-\sum_{n}P_{n}\ln{P_{n}}.
\end{equation}
Here $P_{n}$ is the probability of having {\em n} produced particles in the final state such that $\sum{P_{n}}=1$ for any phase space intervals. The parameter entropy is related to the fractal dimension, to be more specific, the information dimension\cite{simak,ghosh1}. The entropy is invariant under an arbitrary distortion of multiplicity scale, in particular, if a sub-sample of particles is chosen such as charged particles. According to the formula (3) we have calculated the entropy and the reduced entropy ($S/<n>$), i.e. the ratio of the entropy to average multiplicity of target evaporated fragments in the forward and backward hemisphere for all of our investigated nucleus-nucleus interaction. Table\,4 presents our results together with the results in Ref.\cite{ghosh}. It is found that the entropy value of the target evaporated fragments emitted in the forward hemisphere is greater than that emitted in backward hemisphere, and the values of entropies remain almost energy independent in the forward and backward hemispheres respectively.

\begin{table}
\begin{center}
Table 4. The values of entropy, reduced entropy and scaled variance of target evaporated fragments in the forward and backward hemispheres for different type of nucleus-AgBr interactions.\\
\begin{small}
\begin{tabular}{lccccccc}\hline
Beam    &\multicolumn{3}{c}{forward hemisphere} &\multicolumn{3}{c}{backward hemisphere} & Reference  \\
                 & $S$           & $S/<n>$       & {\em w}       & $S$           & $S/<n>$       & $w$           &    \\\hline
12 A GeV He   & $2.38\pm0.12$ & $0.52\pm0.03$ & $1.76\pm0.02$ & $2.20\pm0.09$ & $0.60\pm0.03$ & $1.50\pm0.01$ & This work\\
3.7 A GeV O   & $2.54\pm0.25$ & $0.42\pm0.04$ & $1.72\pm0.02$ & $2.29\pm0.18$ & $0.51\pm0.04$ & $1.39\pm0.01$ & This work\\
60 A GeV O    & $2.36\pm0.27$ & $0.39\pm0.05$ & $1.32\pm0.04$ & $2.28\pm0.23$ & $0.45\pm0.04$ & $1.15\pm0.01$ & This work\\
1.7 A GeV Kr  & $2.49\pm0.44$ & $0.39\pm0.07$ & $1.51\pm0.02$ & $2.26\pm0.37$ & $0.54\pm0.08$ & $1.52\pm0.07$ & This work\\
10.7 A GeV Au & $2.38\pm0.23$ & $0.47\pm0.05$ & $1.46\pm0.01$ & $2.17\pm0.20$ & $0.60\pm0.06$ & $1.41\pm0.02$ & This work\\
3.7 A GeV C   & $2.54\pm0.11$ & $0.39\pm0.01$ & $2.12\pm0.03$ & $2.30\pm0.20$ & $0.43\pm0.01$ & $1.40\pm0.02$ & [3] \\
14.5 A GeV Si & $2.24\pm0.14$ & $0.38\pm0.02$ & $1.15\pm0.01$ & $2.18\pm0.19$ & $0.44\pm0.02$ & $1.12\pm0.02$ & [3]\\
60 A GeV O    & $2.32\pm0.14$ & $0.35\pm0.01$ & $1.25\pm0.02$ & $2.21\pm0.11$ & $0.43\pm0.01$ & $1.25\pm0.03$ & [3] \\
200 A GeV S   & $2.31\pm0.16$ & $0.43\pm0.03$ & $1.60\pm0.05$ & $2.03\pm0.12$ & $0.45\pm0.01$ & $1.35\pm0.03$ & [3]\\\hline
\end{tabular}
\end{small}
\end{center}
\end{table}

Finally we want to evaluate the multiplicity correlation among the target evaporated fragments in the forward and backward hemispheres respectively. A useful measure of the fluctuation of any variable is the ratio of its variance to its mean value. So the variance of the multiplicity distribution can be used to measure the multiplicity fluctuations. In order to study the multiplicity fluctuations in high energy nucleus-nucleus collisions, we use a scaled variable {\em w} suggested by Ghosh et al\cite{ghosh} such that
\begin{equation}
w=\frac{<n^{2}>-<n>^{2}}{<n>}.
\end{equation}

Multiplicity fluctuation is one aspect of a two-particle correlation function. The study of the scaled variance can very easily reveal the nature of the correlation among the produced particles. If the value of {\em w} is much greater than $1$, it may be said that there is a strong correlation among the produced particles. In contrast, if the value of {\em w} is close to one, feeble correlation is indicated. The scaled variance {\em w} of target evaporated fragments emitted in forward and backward hemispheres for our investigated high-energy nucleus-nucleus collisions is presented in Table 4, the corresponding values in Ref.\cite{ghosh} is also shown in the table. It can be seen that the scaled variances of target evaporated fragments are close to one in both hemispheres for all the interactions, which suggested that there is a feeble correlation in production of the target evaporated fragments in forward and backward hemispheres.

\section{Conclusions}
The multiplicity distribution, multiplicity moment, scaled variance, entropy and reduced entropy of target evaporated fragment emitted in forward and backward hemispheres in 12 A GeV $^{4}$He, 3.7 A GeV $^{16}$O, 60 A GeV $^{16}$O, 1.7 A GeV $^{84}$Kr and 10.7 A GeV $^{197}$Au induced emulsion heavy targets (AgBr) interactions are investigated. It is found that the multiplicity distribution of target evaporated fragment emitted in forward and backward hemispheres can be fitted by a Gaussian distribution, the Gaussian fitting parameters are different between the forward and backward hemispheres for all the interactions which may be commented that the nature of the emission of target evaporated particles differs between the two hemispheres. The multiplicity moments of target evaporated particles emitted in forward and backward hemispheres increase with the order of the moment {\em q}, and second-order multiplicity moment of the target evaporated fragment in both hemispheres are energy independent over the entire energy for all the interactions. The scaled variance, a direct measure of multiplicity fluctuations, is close to one for all the interactions which may be said that there is a feeble correlation among the target evaporated fragments. The entropy of target evaporated fragment emitted in forward and backward hemispheres are the same within experimental errors, respectively.

\section{Acknowledgements}
This work has been supported by the Chinese National Science Foundation under Grant No: 11075100 and the Natural Foundation of Shanxi Province under Grant 2011011001-2, the Shanxi Provincial Foundation for Returned Overseas Chinese Scholars, China (Grant No. 2011-058). We are grateful to EMU-01 collaboration for supplying emulsion stacks.

\end{document}